\documentclass[amssymb,prb,twocolumn,showpacs,superscriptaddress]{revtex4}
\usepackage{epsfig}
\usepackage{dcolumn}
\usepackage{amsmath,amssymb}
\let \Re \relax

\def \beq {\begin{equation}}
\def \edq {\end{equation}}
\def \bes {\begin{subequations}}
\def \eds {\end{subequations}}
\def \beqn {\begin{equation*}}
\def \edqn {\end{equation*}}
\def \nn  {\nonumber}
\def \dag {\dagger}

\def \up {\uparrow}
\def \down {\downarrow}

\def \sm {\sigma}
\def \bsm {\bar{\sigma}}

\def \veps {\varepsilon}
\def \cala {{\cal{A}}}

\def \calh {{\cal{H}}}

\def \calg {{\cal{G}}}

\def \calt {{\cal{T}}}

\def \wteps {\widetilde{\veps}}

\def \wG {\widetilde{\Gamma}}

\def \wD {\widetilde{D}}
\def \trho {\widetilde{\rho}}
\def \wPsi {\widetilde{\Psi}}

\providecommand{\nbraket}[1]{\langle#1\rangle}
\providecommand{\dbraket}[1]{\langle\langle#1\rangle\rangle}

\DeclareMathOperator{\Re}{Re}

\begin{document}

\title{Kondo effect in spin-orbit mesoscopic interferometers}
\author{Jong Soo Lim}
\affiliation{Departament de F\'{\i}sica,
Universitat de les Illes Balears, E-07122 Palma de Mallorca, Spain}
\author{Mircea Crisan}
\affiliation{Department of Theoretical Physics, University of Cluj,
3400 Cluj, Romania}
\author{David S\'anchez}
\affiliation{Departament de F\'{\i}sica,
Universitat de les Illes Balears, E-07122 Palma de Mallorca, Spain}
\affiliation{Institut de F\'{\i}sica Interdisciplinar i de Sistemes Complexos
IFISC (CSIC-UIB), E-07122 Palma de Mallorca, Spain}
\author{Rosa L\'opez}
\affiliation{Departament de F\'{\i}sica,
Universitat de les Illes Balears, E-07122 Palma de Mallorca, Spain}
\affiliation{Institut de F\'{\i}sica Interdisciplinar i de Sistemes Complexos
IFISC (CSIC-UIB), E-07122 Palma de Mallorca, Spain}
\author{Ioan Grosu}
\affiliation{Department of Theoretical Physics, University of Cluj,
3400 Cluj, Romania}

\date{\today}

\begin{abstract}
We consider a flux-threaded Aharonov-Bohm ring with an embedded quantum dot
coupled to two normal leads. 
The local Rashba spin-orbit interaction acting on the dot electrons leads
to a spin-dependent phase factor
in addition to the Aharonov-Bohm phase caused by the external flux.
Using the numerical renormalization group method,
we find a splitting of the Kondo resonance at the Fermi level
which can be compensated by an external magnetic field.
To fully understand the nature of this compensation effect,
we perform a scaling analysis and derive an expression for the effective magnetic field.
The analysis is based on a tight-binding model which leads to an effective Anderson model
with a spin-dependent density of states for the transformed lead states.
We find that the effective field originates from the combined effect of Rashba interaction
and magnetic flux and that it contains important
corrections due to electron-electron interactions.
We show that the compensating field
is an oscillatory function of both the spin-orbit and the Aharonov-Bohm phases.
Moreover, the effective field never vanishes due to the particle-hole symmetry breaking
independently of the gate voltage.
\end{abstract}


\maketitle

\section{Introduction}\label{sec_intro}
Interference studies in solid-state mesoscopic interferometers
provide most valuable information about scattering properties
of artificial atoms (quantum dots).\cite{yac95,lev95,hac96,bru96,shu97,ji00,sig04}
Interference takes place between electronic partial waves
traveling along a nonresonant channel (the reference arm)
and through a quasi-localized state (the quantum dot).
When an external magnetic flux is piercing the area enclosed
by the interferometer, the partial waves pick up different
Aharonov-Bohm phases and conductance oscillations are observed.
Moreover, path interaction between the background channel and
hopping through the dot gives rise to characteristic asymmetric Fano
transmission lineshapes.\cite{kob02}

For strongly interacting dots which are coupled to external reservoirs,
transport at low temperatures is dominated by Kondo correlations,
which originate from a nontrivial antiferromagnetic interaction
between the leads' conduction electrons and the dot electron
in a discrete level playing the role of a quantum impurity.\cite{Hewson93}
Such interaction leads to a screening of the impurity spin
and the linear conductance reaches in the strong coupling regime
the maximum value $2e^ 2/h$ (the unitary limit) for a wide range
of the gate voltage.\cite{Kondo}
However, when the dot is inserted in the arm
of an Aharonov-Bohm interferometer, 
the linear-response curves evolve from the unitary limit
to asymmetric lineshapes and finally
to a plateau of zero conductance as the background
transmission $T_b$ increases from 0 to 1.\cite{hof01}
In addition, the differential conductance shows a zero-bias
peak at $T_b=0$ which is transformed into a dip when
$T_b$ approaches 1.\cite{bul01}

At the same time, spin-orbit interactions have been
a subject of ongoing interest since the advent of spintronics.\cite{fab07}
A prominent spin-orbit interaction is the Rashba interaction, which
arises in inversion asymmetric semiconductor heterostructures.\cite{Rashba60}
Aharonov-Bohm oscillations have been observed in rings in the presence
of spin-orbit interactions.\cite{morpurgo,yau,yang,grbic}
When a dot subject to Rashba interaction is embedded in the mesoscopic interferometer,
the traveling electrons acquire a spin-dependent phase in addition to the
Aharonov-Bohm phase.\cite{sun05} A similar effect takes place
in quantum wires with localized Rashba coupling,\cite{lop07} in which case
localized magnetic states can be formed in nonequilibrium situations.\cite{cri09} 
For spin-orbit quantum-dot Aharonov-Bohm systems,
the spin polarization can be controlled by tuning the magnetic flux
and the Rashba strength.\cite{hea08} More importantly, using numerical
renormalization group methods it has been
argued that a compensation effect takes place when an external
Zeeman field is applied, eliminating the splitting due to the spin-orbit interaction
in the Kondo local density of states.\cite{ver09}
This situation is also seen in single dots coupled to ferromagnetic cases,
for which spin dependent coupling leads to the splitting of the dot spectral
weights\cite{martinek} via an effective field
for gate voltages away from the particle-hole symmetric point.\cite{choi}

The nature of the compensating field can be fully understood
only through a scaling analysis. This is the goal we want to accomplish
in the present work. We find that the origin of the effective field
is twofold: (i) for noninteracting electrons the combination
of spin-orbit interactions and external flux gives rise to a splitting
of the dot energy levels; and (ii) in the presence of interactions,
described only beyond mean-field theory, the effective field acquires
corrections which are of the same order as the noninteracting value
in the case of very strong correlations. We find that
the compensating field is a {\em periodic} function of the spin-orbit
and the Aharonov-Bohm phases and that it is always nonzero {\em independently}
of the gate voltage due to the breaking of the particle-hole symmetry point.
Our results are complemented with a mean-field
theoretical approach and perturbation theory, in agreement
with the exact numerical calculations.

The paper is organized as follows.
In Sec.~\ref{sec:model}, we introduce a theoretical model and review the noninteracting solution.
We calculate the effective field that splits the dot level including interactions
at the mean field level.
We then show in Sec.~\ref{sec:nrg} our results from numerical renormalization group calculations,
obtaining a splitting of the Kondo peak.
To reveal the origin of this splitting, we perform in Sec.~\ref{sec:scaling}
a two-stage scaling analysis and obtain the renormalized
dot levels in the presence of both the Aharonov-Bohm and the Rashba phases.
In Sec.~\ref{sec:Beff}, we derive an explicit expression for the effective magnetic
field using a Schrieffer--Wolff-like mean-field theory. Finally, our conclusions
are contained in Sec.~\ref{sec:concl}.

\section{Theoretical model} \label{sec:model}

\begin{figure}
\centering
\includegraphics[width=0.4\textwidth]{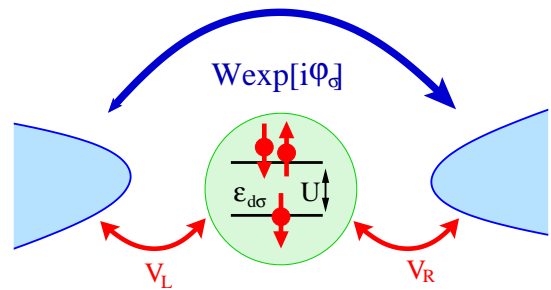}
\caption{Sketch of the system under consideration. A two-lead Aharonov-Bohm interferometer
has a quantum dot embedded in its lower arm. $W$ is the transmission amplitude
in the direct path whereas $V_\alpha$ ($\alpha = L/R$) are hopping matrix
elements between the dot and the leads. The electrons traveling along the ring can
acquire a spin-dependent phase $\phi_\sigma$. We consider a quantum dot with a single energy
level $\varepsilon_d$ and charging energy $U$.
}
\label{figsystem}
\end{figure}

We consider an Aharonov-Bohm (AB) inteferometer
in contact with two normal leads, see Fig.~\ref{figsystem}.
A quantum dot with local Rashba spin-orbit interaction
is embedded in one of two arms of the AB interferometer.
The Hamiltonian of the system under consideration is then,
\beq
\calh = \calh_{D} + \calh_C + \calh_T \,,
\label{eq:model}
\edq
where
\bes
\begin{align}
\calh_D &= \sum_{\sm} \veps_{d\sm} d_{\sm}^{\dag}d_{\sm} + Un_{d\up}n_{d\down} \,,\\
\calh_C &= \sum_{\alpha = L/R,k,\sm} \veps_{k\sm} c_{\alpha k\sm}^{\dag}c_{\alpha k\sm} \,, \\
\calh_T &= \sum_{\alpha,k,\sm} \left[ V_{\alpha} d_{\sm}^{\dag}c_{\alpha k\sm} + h.c.\right] \nn \\
&+ \sum_{k,k',\sm} \left[W e^{i\phi_{\sm}} c_{Lk\sm}^{\dag}c_{Rk'\sm} + h.c.\right] \,.
\end{align}
\eds
Here, $\calh_D$ describes the quantum dot Hamiltonian with single-particle energy
$\epsilon_{d\sm}$ and on-site Coulomb repulsion $U$.
$n_{\sm}=d_{\sm}^{\dag}d_{\sm}$ denotes the occupation of the dot. 
$\calh_C$ represents two normal leads ($\alpha = L/R$)
and the tunneling between leads and AB ring is given by $\calh_T$.
In the tunneling Hamiltonian, the coupling $V_{\alpha}$  describes the electron
tunneling between lead $\alpha$ and the quantum dot,
while $W$ corresponds to the direct tunneling amplitude between the leads.
Due to the presence of a flux $\Phi$ threading the area enclosed by the ring,
the electron acquires the AB phase  $\phi_{AB} = 2\pi\Phi/\Phi_0$, where $(\Phi_0 = hc/e)$
is the flux quantum.\cite{zeeman} At the same time, the dot is subject to spin-orbit interactions which
give rise to the {\em spin-dependent} phase $\sm\phi_{SO}$. \cite{sun05} As a consequence,
the total phase accumulated along the ring is $\phi_{\sm} = \phi_{AB} + \sm\phi_{SO}$.

\subsection{Noninteracting case}
We first review the non-interacting case ($U=0$).\cite{ver09} We define
the quantum-dot retarded Green function for electrons with spin $\sigma$,
\beq
\calg_{d\sm}^r(\omega)=\int_{-\infty}^\infty dt\, e^{i \omega t} \dbraket{d_{\sm}(t),d_{\sm}^\dagger(0)}^r\,,
\edq
where $\dbraket{d_{\sm}(t),d_{\sm}^\dagger(0)}^r = -i\Theta(t) \nbraket{[d_{\sm}(t),d_{\sm}^\dagger(0)]_+}$.
The Green function can be calculated exactly as,
\beq\label{eq_gdsm}
\calg_{d\sm}(\omega)  = \frac{1}{\omega - \veps_{d\sm} - \Sigma_{0\sm}} \,,
\edq
where the zeroth-order (i.e., noninteracting) self-energy reads,
\beq
\Sigma_{0\sm} = \frac{\cala(V_L^2 + V_R^2 + 2V_LV_R W\cos(\phi_{\sm})\cala)}{1 - W^2\cala^2} \,,
\label{eq:selfenergy}
\edq
with
\beq
\cala(z \to \omega + i0^+) \equiv \sum_k \frac{1}{z-\veps_{k\sm}} \,.
\edq

In the wide-band limit, the self-energy can be written as
\beq
\Sigma_{0\sm} = -\wG \left(i + \sqrt{\alpha\xi}\cos(\phi_{\sm})\right) \,,
\label{eq:self0}
\edq
with $\rho_0 = \rho(E_F)$, $\xi = \pi^2\rho_0^2 W^2$, $\alpha = 4V_L^2V_R^2/(V_L^2+V_R^2)^2$, and $\wG \equiv \Gamma/(1+\xi) = \pi\rho_0 (V_L^2+V_R^2)/(1 + \xi)$. From Eq.~\eqref{eq:self0}, we note that the real part of the self-energy,
$\Re\left[\Sigma_{0\sm}\right]$, is spin-dependent because of the factor $\cos\phi_{\sm}$.
In turn, this implies that there is an {\em effective} Zeeman field,
\beq
g\mu_B B_{\rm eff} = \wG\sqrt{\alpha\xi} \sin\phi_{AB}\sin\phi_{\rm SO}\,,\label{eq:nicfield}
\edq
acting on the dot levels due to the combined effect of the local Rashba interaction
and the magnetic flux. This effect appears because
the phase acquired by an electron with a given spin orientation
is $\phi_{AB}+\phi_{SO}$ whereas the opposite orientation acquires $\phi_{AB}-\phi_{SO}$.
In fact, if either $\phi_{AB}$ or $\phi_{\rm SO}$ vanish, we have $B_{\rm eff}=0$.
Then, quite generally, $B_{\rm eff}$ can lead to net spin polarizations
in the transmitted current.\cite{sun05,hea08} 

Trivially, such effective field can be canceled by an externally applied field $B_{\rm ext}$
such that $B_{\rm ext} = -B_{\rm eff}\equiv B_c$ which compensates the splitting.
We see that the compensation field $B_c$ is a periodic function of $\phi_{AB}$ and $\phi_{\rm SO}$.
Our next goal is to include interactions.

\subsection{Mean-field approximation}
In the simplest approach that includes interaction,\cite{anderson61} one replaces $\veps_{d\sm}$
with $E_{d\sigma}=\veps_{d\sm}+U\langle n_{\bsm}\rangle$ in Eq.~\eqref{eq_gdsm},
where the mean dot occupation at equilibrium reads,
\beq\label{eq:nsm:new}
\langle n_\sm\rangle=-\frac{1}{\pi}\int_{-\infty}^\infty d\omega \, f(\omega)
\,{\rm Im}\,\calg_{d\sm}^r(\omega)\,
\edq
with $f(\omega)$ the Fermi distribution function. The problem, thus, must be self-consistently
solved since the right-hand side of Eq.~\eqref{eq:nsm:new} depends on $\langle n_\sm\rangle$.
This Hartree approach is known to generate local magnetic moments,
even in the presence of spin-orbit interactions.\cite{cri09}
To avoid this, we will here focus on the nonmagnetic phase.

We extend the method of Ref.~\onlinecite{hor} to account for
both the AB and the spin-orbit phases. Then,
we obtain for the special point $\varepsilon_d=-U/2$
the self-consistent equation,
\beq
E_{d\sigma}=-\sigma\frac{U}{\pi}\tan^{-1} \left(\frac{U m}{2\wG}\right)
-\wG \sqrt{\alpha\xi}\cos(\phi_{\sm})\,,
\edq
Here, $m=\langle n_\uparrow\rangle-\langle n_\downarrow\rangle$
is the dot magnetization. The compensating field is calculated from
the condition $m=0$, which is satisfied by
\beq
g\mu_B B_{c}=-\wG\sqrt{\alpha\xi} \sin\phi_{AB}\sin\phi_{\rm SO}\,.\label{eq:bcmf}
\edq
We note this value coincides with the noninteracting result obtained above.
Therefore, to find corrections to the noninteracting case we must go {\em beyond}
the mean-field approach and include strong correlations.
We first analyze the problem numerically and then later we perform
a scaling study that demonstrates that there are indeed corrections
due to interactions but, strikingly, the periodicity of $B_c$ is preserved.

\section{Numerical Renormalization Group Calculation} \label{sec:nrg}

We now present a numerical renormalization group (NRG) analysis of our system.
Employing the standard NRG recipe \cite{Kri80} and an even/odd parity basis,
\bes
\begin{align}
f_{ne\sigma} &= \frac{1}{\sqrt{2}} \left( e^{-i\phi_{\sm}/2}f_{nL\sigma} + e^{+i\phi_{\sm}/2}f_{nR\sigma} \right) \,,\\
f_{no\sigma} &= \frac{1}{\sqrt{2}} \left( ie^{-i\phi_{\sm}/2}f_{nL\sigma} - ie^{+i\phi_{\sm}/2}f_{nR\sigma} \right) \,,
\end{align}
\eds
we first map the continuous conduction bands onto the corresponding tight-binding model.
Here, a symmetric coupling is taken, i.e., $V_L = V_R \equiv V_1$.
The resulting Hamiltonian can be then written as
\begin{multline}
\calh =  \sum_{\sigma} \sqrt{\frac{2\Gamma}{\pi}} \left( \cos(\phi_{\sm}/2) d_{\sigma}^{\dag}f_{0e\sigma} + \sin(\phi_{\sm}/2) d_{\sigma}^{\dag}f_{0o\sigma} + h.c. \right) \\
+ \frac{1+\Lambda^{-1}}{2} \sum_{\alpha=e/o}\sum_{n=0}^{\infty}\sum_{\sigma} \Lambda^{-n/2} \zeta_n \left( f_{n\alpha\sigma}^{\dag}f_{n+1\alpha\sigma} + h.c.\right) \\
+ \frac{2}{\pi} \sqrt{\xi}\sum_{\sigma} \left(f_{0e\sigma}^{\dag}f_{0e\sigma} - f_{0o\sigma}^{\dag}f_{0o\sigma} \right) + \calh_D  \,.
\end{multline}
Here, we assume a constant conduction band with a half-width $D$.
Note that the couplings between lead and dot are now spin-dependent.
To solve the Hamiltonian, we define a sequence of Hamiltonians $\bar{\calh}_N$ as follows:
\begin{multline}
\bar{\calh}_N = \Lambda^{(N-1)/2} \left\{ \calh_{D} \right. \\
+ \sum_{\sigma} \sqrt{\frac{2\Gamma}{\pi}} \left( \cos(\phi_{\sm}/2) d_{\sigma}^{\dag}f_{0e\sigma} + \sin(\phi_{\sm}/2) d_{\sigma}^{\dag}f_{0o\sigma} + h.c. \right)  \\
+ \frac{1+\Lambda^{-1}}{2} \sum_{\alpha=e/o}\sum_{n=0}^{N-1}\sum_{\sigma} \Lambda^{-n/2} \zeta_n \left( f_{n\alpha\sigma}^{\dag}f_{n+1\alpha\sigma} + h.c.\right)  \\
\left. + \frac{2}{\pi} \sqrt{\xi} \sum_{\sigma} \left(f_{0e\sigma}^{\dag}f_{0e\sigma} - f_{0o\sigma}^{\dag}f_{0o\sigma} \right) \right\} \,,
\end{multline}
that results in the recursion relation below
\beq
\widetilde{\calh}_{N+1} = \sqrt{\Lambda} \widetilde{\calh}_{N} 
+ \sum_{\alpha=e/o}\sum_{\sigma} \zeta_n \left( f_{N\alpha\sigma}^{\dag}f_{N+1\alpha\sigma} + h.c.\right) \,,
\edq
with
\beq
\widetilde{\calh}_{N} = \frac{2}{1+\Lambda^{-1}}\bar{\calh}_{N} \,.
\edq
Using this recursion relation, we iteratively diagonalize the NRG Hamiltonian and keep only the lowest eigenvalues in each step.
In doing so, however, we have to be careful because the original Wilson's NRG approach\cite{wilson}
fails in the presence of a magnetic field.
This failure results from the fact that at the initial stages
the Wilson's approach does not yet know about the tiny perturbation breaking the spin symmetry
and thus yields an incorrect ground state.
Here, we thus employ the so-called density-matrix (DM)
NRG approach developed by Hofstetter.\cite{Hof00}
Although there exist more sophisticated methods in the literature,\cite{Wei07}
we believe the DM-NRG is enough for our purposes.

\begin{figure}
\centering
\includegraphics[width=0.4\textwidth]{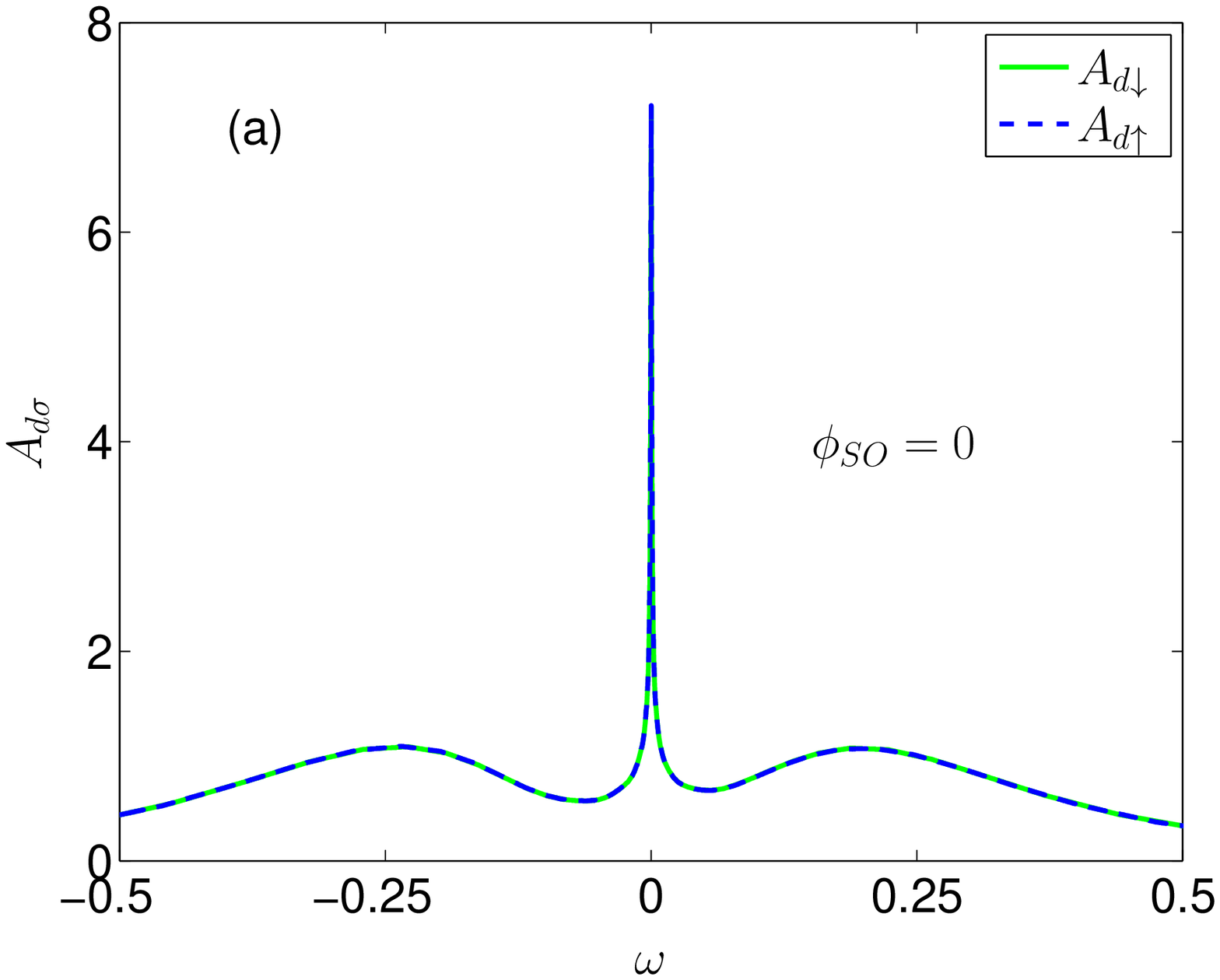} \\
\includegraphics[width=0.4\textwidth]{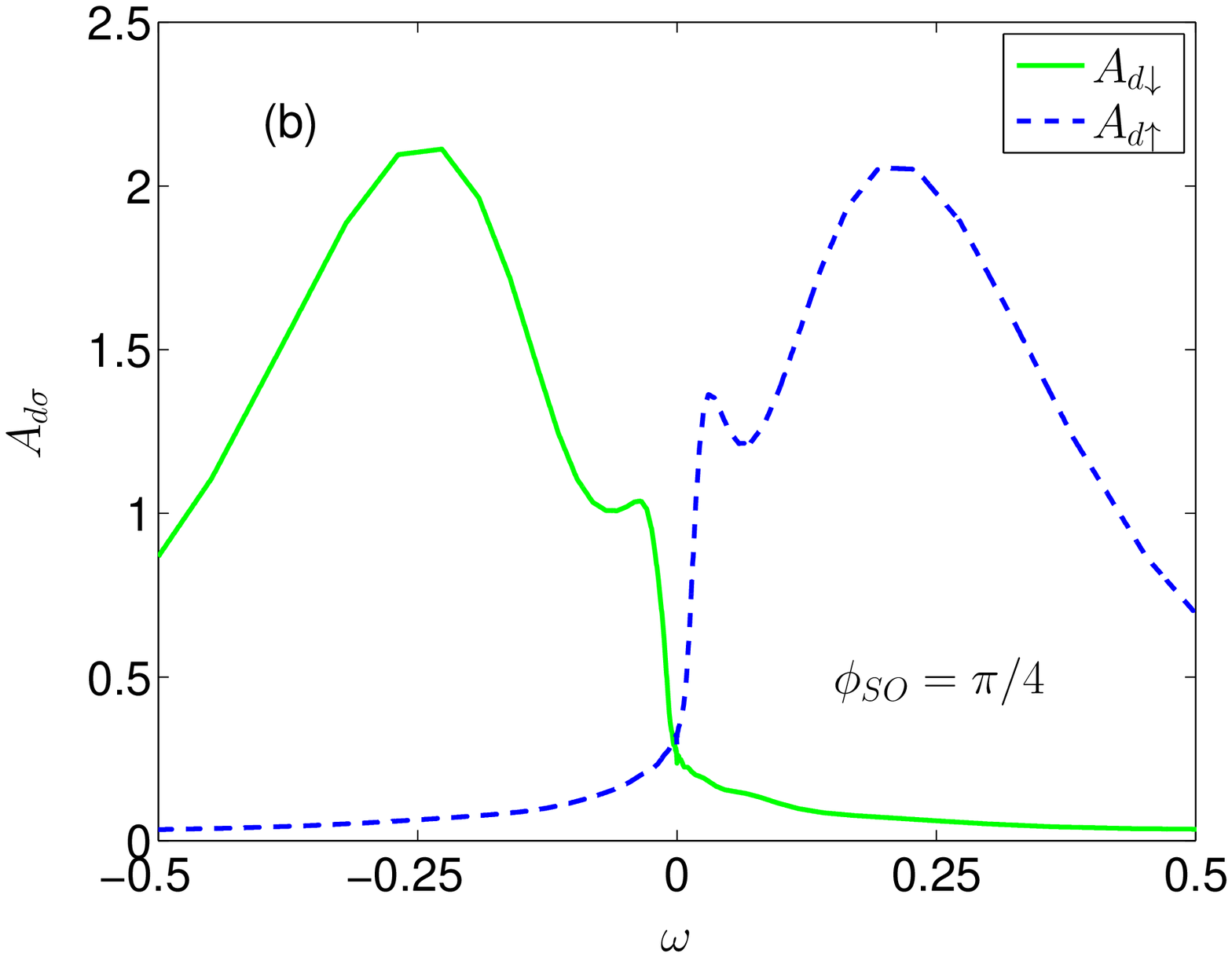}
\caption[Spectral weights]{Spectral weights. Parameters are $D=1$, $\veps_d = -0.25$, $U=0.5$, $V_1 = 0.1414$, $\xi = 1$, $\phi_{AB} = \pi/4$.
The spin-orbit coupling strengths are (a) $\phi_{SO} = 0$ and (b) $\pi/4$.} 
\label{fig:splitting}
\end{figure}

\begin{figure}
\centering
\includegraphics[width=0.4\textwidth]{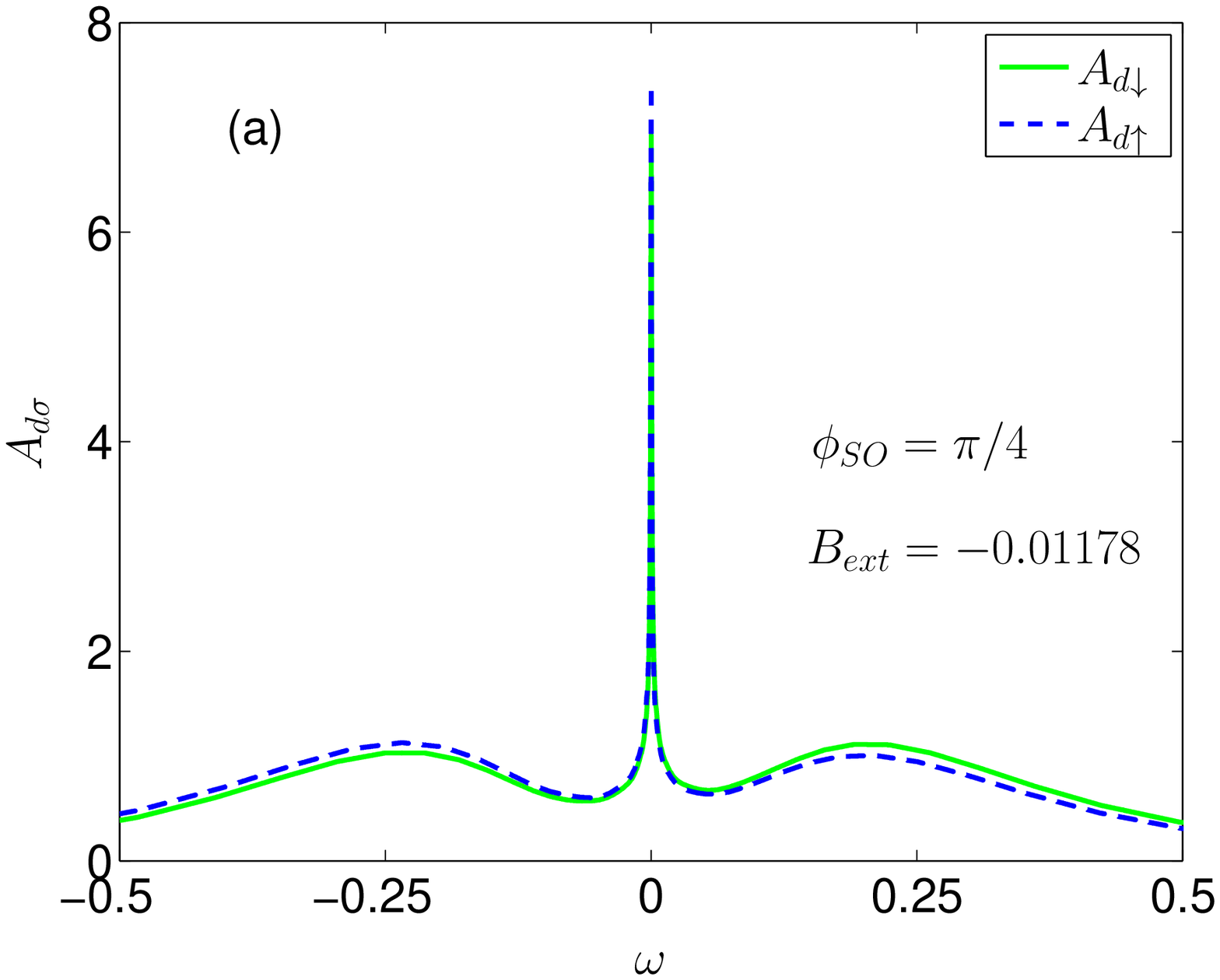} \\
\includegraphics[width=0.4\textwidth]{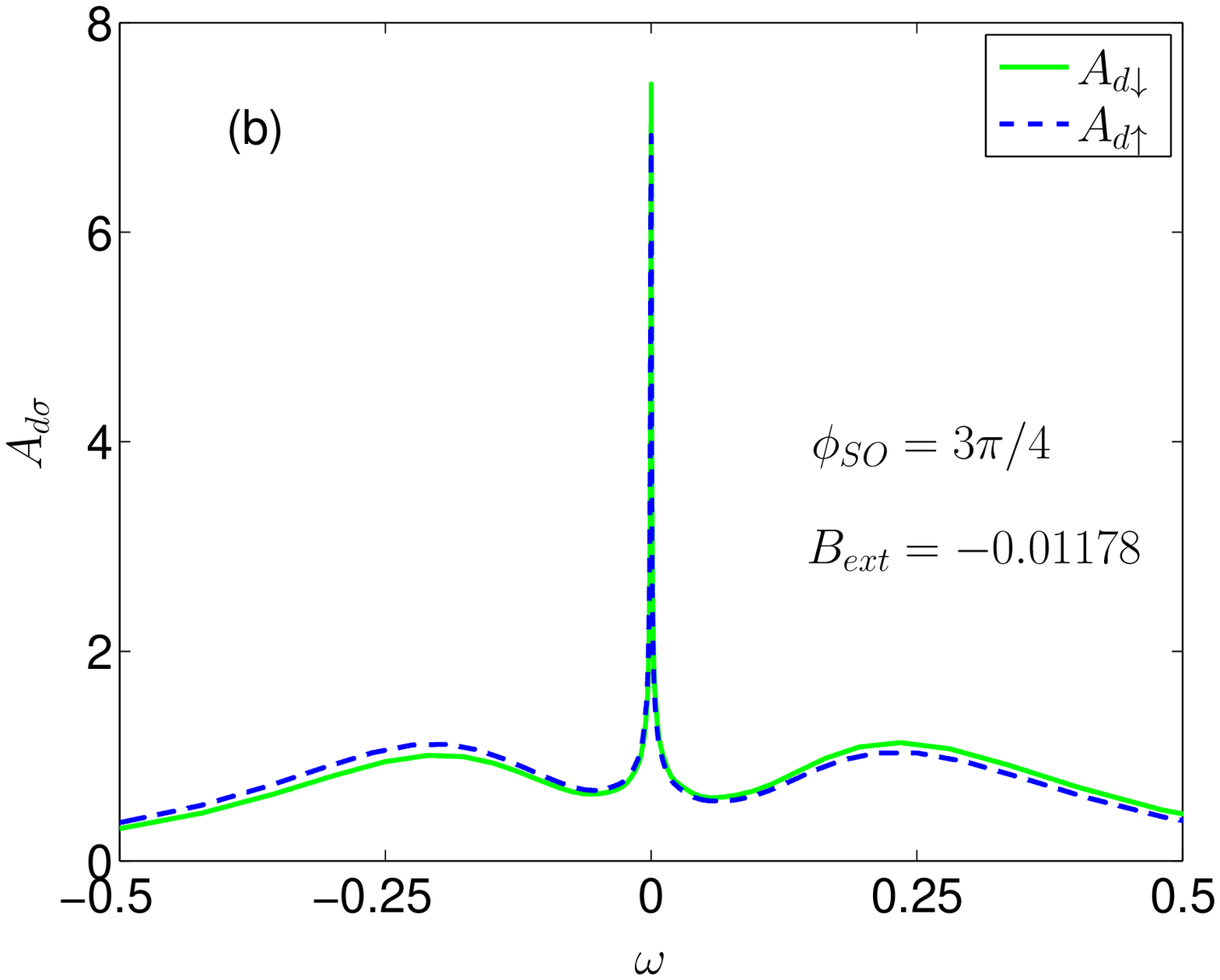} \\
\includegraphics[width=0.4\textwidth]{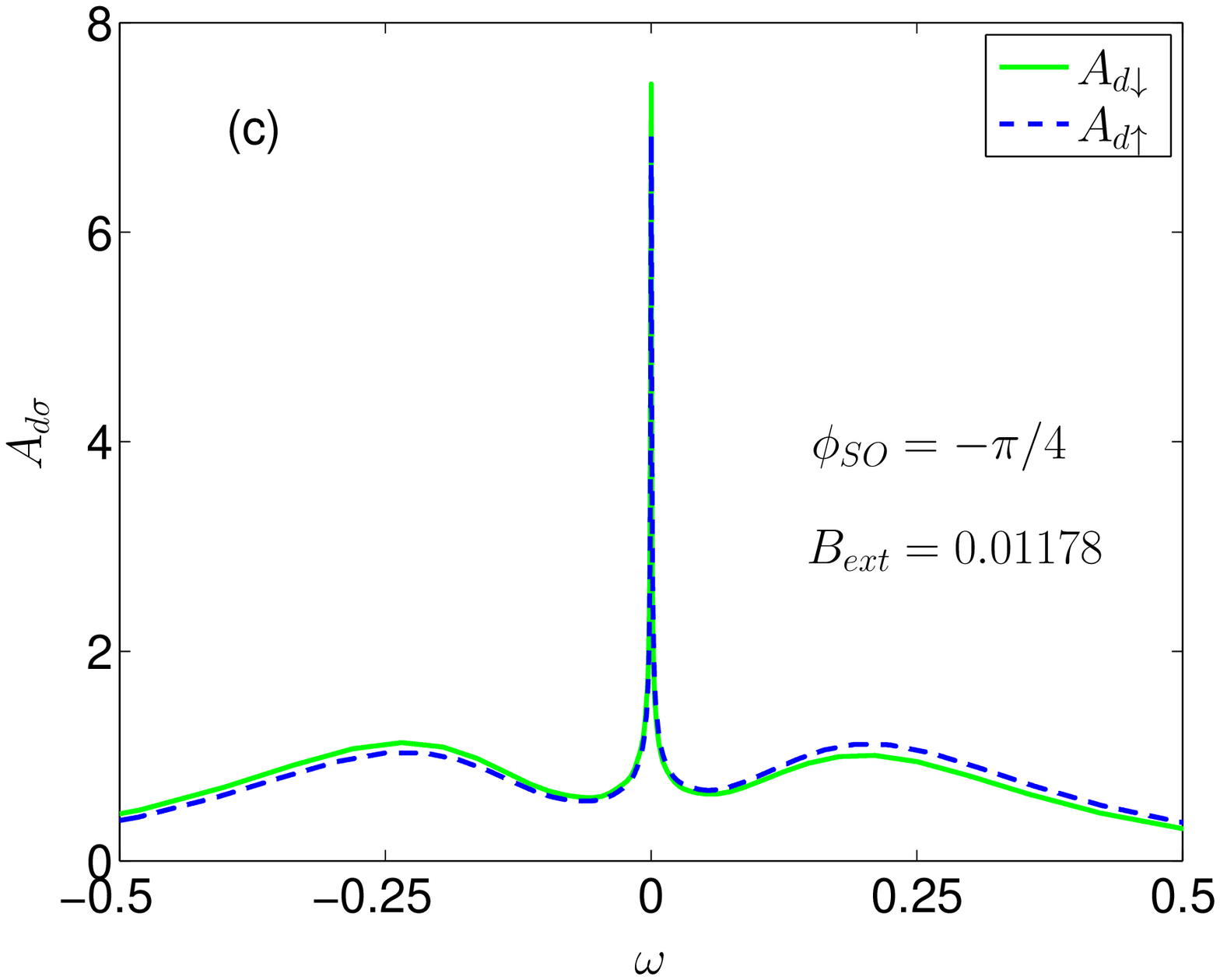} \\
\includegraphics[width=0.4\textwidth]{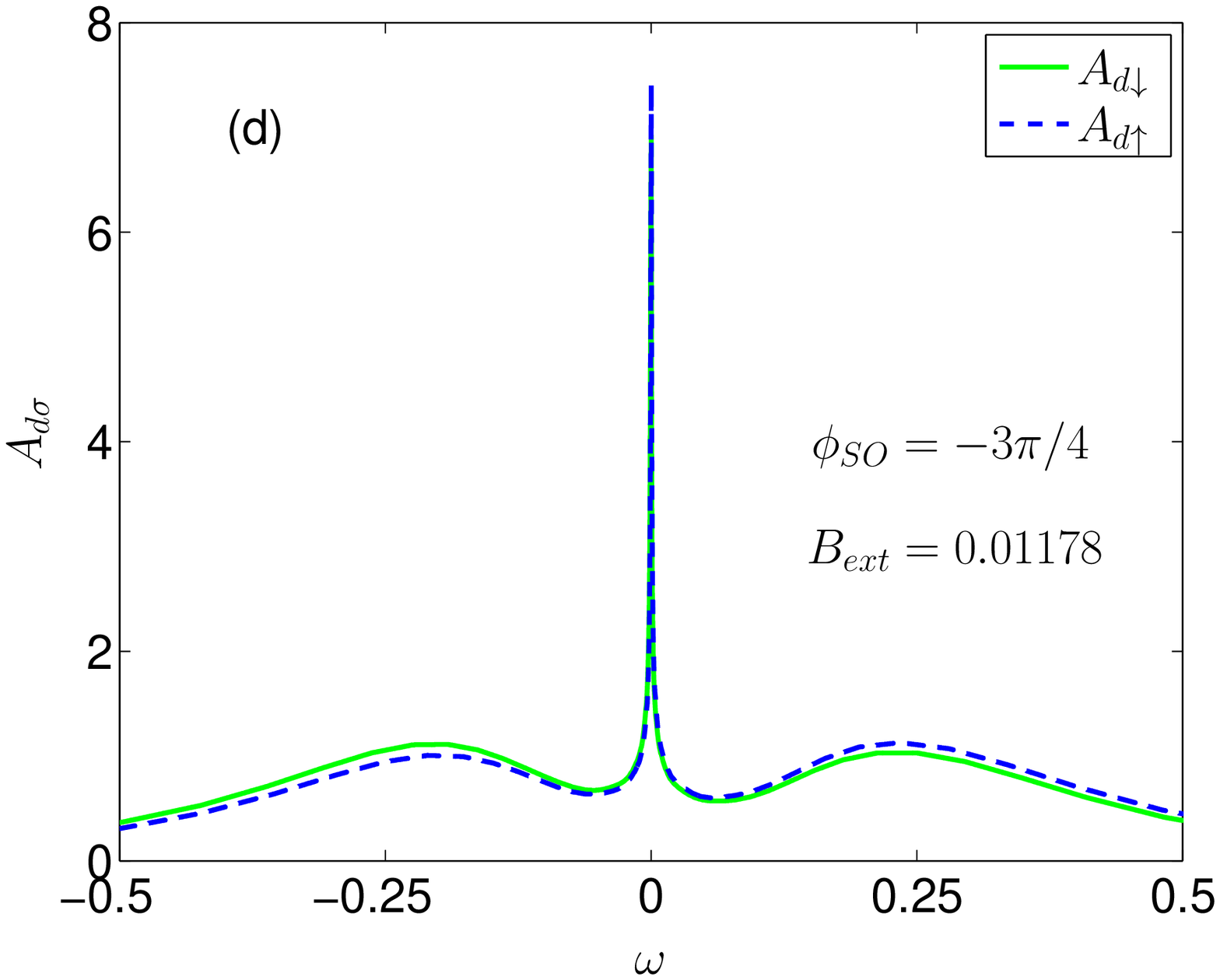}
\caption[Compensation effect]
{Compensation effect. The spin-orbit strengths are (a) $\phi_{SO} = \pi/4$, (b) $\phi_{SO} = 3\pi/4$, and (c) $\phi_{SO} = -\pi/4$, and (d) $\phi_{SO} = -3\pi/4$.
Parameters are $D=1$, $\veps_d = -0.25$, $U=0.5$, $V_1 = 0.1414$, $\xi = 1$, and $\phi_{AB} = \pi/4$.} 
\label{fig:compensation}
\end{figure}

In Fig.~\ref{fig:splitting}, we show for both spin up and spin down
the spectral weights of the dot as a function of the Rashba spin-orbit interaction $\phi_{SO}$
for $\phi_{AB}=\pi/4$.
Without spin-orbit interactions [Fig.~\ref{fig:splitting}(a)],
both spectral weights coincide and do not split.
However, for $\phi_{SO} = \pi/4$ [Fig.~\ref{fig:splitting}(b)] the weights split
and the Kondo resonances near the Fermi level become suppressed.\cite{ver09}
Moreover, the spectral weights move to the particle (hole) sector for spin up (down).
Such a shift results from the polarization of the dot occupation.

The split Kondo peaks can be compensated by applying an external magnetic field.
Figure~\ref{fig:compensation} shows the recovery of the Kondo resonance
at the Fermi level for various values of $\phi_{SO}$.
We observe that the compensating fields for $\phi_{SO} = \pi/4 (3\pi/4)$ and $-\pi/4 (-3\pi/4)$
have opposite signs. Therefore, the effective field is invariant under simultaneous
reversal of both the AB flux and the Rashba interaction. This fact is understood
in the noninteracting case from Eq.~\eqref{eq:nicfield}. Hence, it is
crucial to investigate in detail the precise form of the compensating field
in the presence of interactions. This goal can be accomplished only
via a scaling analysis, which we perform in the next section.

\section{Scaling Analysis} \label{sec:scaling}

\begin{figure}
\centering
\includegraphics[width=0.45\textwidth]{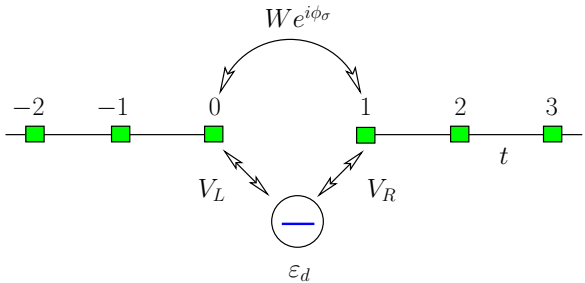}
\caption{AB ring with spin-orbit interaction of the Rashba type and coupled to semi-infinite discrete leads.}
\label{fig:tightbindingso}
\end{figure}

\subsection{Tight-Binding Model and Lead Polarization}

The scaling analysis is greatly simplified if we consider a tight-binding
model of the system. 
We discretize the leads and consider
the system shown in Fig.~\ref{fig:tightbindingso}.
The model Hamiltonian is given by
\beq
\calh = \calh_{D} + \calh_C + \calh_T \,,
\label{eq:absotb}
\edq
where
\bes
\begin{align}
\calh_D &= \sum_{\sm} \veps_{d\sm}d_{\sm}^{\dag}d_{\sm} + Un_{d\up}n_{d\down} \,, \\
\calh_C &= \sum_{n\ne 0}\sum_{\sm} \left(-t c_{n+1\sm}^{\dag}c_{n\sm} + h.c.\right) \nonumber \\
&+ \sum_{\sm}\left(We^{i\phi_{\sm}}c_{0,\sm}^{\dag}c_{1,\sm} + h.c.\right) \,, \\
\calh_T &= \sum_{\sm} \left(V_L d_{\sm}^{\dag}c_{0,\sm} + V_Rd_{\sm}^{\dag}c_{1,\sm} + h.c.\right) \,.
\end{align}
\eds
The only difference with our starting Hamiltonian Eq.~\eqref{eq:model} is
the distinct representation of the leads but this does not change the underlying physics.

The Hamiltonian Eq.~\eqref{eq:absotb} has been considered in Ref.~\onlinecite{yos08}
for the case without spin-orbit interactions.
It is shown that an AB ring with an embedded quantum dot threaded by $\phi_{AB}$
can be mapped onto an equivalent Anderson model
in which the density of states in the transformed lead has a term proportional
to $\veps_{k\sm}\cos(\phi_{AB})$ in addition to a constant.\cite{yos08}
Following their approach, we first diagonalize the lead Hamiltonian $\calh_C$.
Using this diagonalized basis and neglecting the decoupled mode,
we find that Eq.~\eqref{eq:absotb} becomes,
\beq
\calh = \calh_D + \sum_{k,\sm} \veps_{k\sm} c_{sk\sm}^{\dag}c_{sk\sm} + \sum_{\sm} V \left[d_{\sm}^{\dag}c_{sk\sm} + h.c.\right] \,,
\label{eq:effham}
\edq
where the effective density of states in the lead reads
\beq
\rho_{\sm}(\veps_{k\sm}) = \trho_0\left(1 + \sqrt{\alpha \calt_b}\cos(\phi_{\sm}) \frac{\veps_{k\sm}}{D_0}\right) \,,
\label{eq:effrhoso}
\edq
with $V = \sqrt{V_L^2 + V_R^2}$, $x = (V_2/t)^2$ (which amounts to $\xi$ in the continuum model of leads, see Sec.~II), $\trho_0 = 1/\pi t(1+x)$, $\alpha = 4V_L^2V_R^2/(V_L^2+V_R^2)^2$, $\calt_b = 4x/(1+x)^2$, and $D_0 = 2t$. 
Remarkably, from Eq.~\eqref{eq:effrhoso} we observe that the density of states
for the reduced lead becomes spin-dependent.
Therefore, we expect a spin-dependent coupling between the lead and the dot which will give rise
to an effective magnetic field in the dot. This situation is also seen in 
simple models of ferromagnetic leads attached to Kondo impurities,\cite{martinek,choi,krawiec}
but the difference is that
while in the latter case the term yielding a spin polarization is constant in energy,\cite{martinek,choi,krawiec}
in our case the density of states contains a term linear in $\veps_{k\sm}$.

To gain further insight into the spin polarized tunnel coupling
arisen from Eq.~\eqref{eq:effrhoso},
it is sufficient to consider the spin-dependent occupation of the
reduced lead at zero temperature,
\beq\label{eq:nsm}
n_{\sm} = \int_{-D_0}^{0} d\veps~ \rho_{\sm}(\veps) \,.
\edq
Defining the effective spin polarization as
\beq
P = \frac{n_{\up} - n_{\down}}{n_{\up} + n_{\down}}\,,
\edq
we insert Eq.~\eqref{eq:effrhoso} in Eq.~\eqref{eq:nsm}
and obtain,
\beq
P = \frac{\sqrt{\alpha\calt_b}\sin(\phi_{AB})\sin(\phi_{SO})}{2 - \sqrt{\alpha\calt_b}\cos(\phi_{AB})\cos(\phi_{SO})}
\,.
\label{eq:polarization}
\edq
We observe that the effective lead polarization
$P$ is zero for $\phi_{AB} = n\pi$ or $\phi_{SO} = n\pi$ with $n$ integer
and depends on the coupling asymmetry $\alpha$ and the background transmission $\calt_b$. 
Moreover, it is worth noting that the fully polarized case ($|P| =1$) can never be achieved
since $|P_{\rm max}|=\sqrt{\alpha T_b}/2 \le 0.5$.

We now calculate the compensating field in the tight-binding representation with $U=0$.
As we know from Sec.~II, the effective field arises from the real part of the tunneling
self-energy which now reads,
\beq\label{eq:resigma}
\Re\Sigma_{0\sm}=V^2\int' d\varepsilon_{k\sm}\,\frac{\rho(\varepsilon_{k\sm})}
{\omega-\varepsilon_{k\sm}}\,,
\edq
where the prime at the integral means the Cauchy's principal value.
The compensating field occurs at external fields such that
$\varepsilon_{d\uparrow}-\varepsilon_{d\downarrow}+\Re(\Sigma_{0\uparrow}-\Sigma_{0\downarrow})=0$
Using Eqs.~\eqref{eq:effrhoso} and~\eqref{eq:resigma}, we find
\beq\label{eq:bcnonint}
g\mu_B B_{\rm c}=-\wG_t\sqrt{\alpha\calt_b} \sin\phi_{AB}\sin\phi_{\rm SO}\,,
\edq
with $\wG_t=2\trho_0V^ 2$.
This results agrees with the
compensating field obtained in the continuum model of Sec.~II,
see Eq.~\eqref{eq:nicfield},
up to a factor $1/(1+x)$ (or $1/(1+\xi)$).
Although the prefactors are different, the functional
dependence is the same.

\subsection{Compensating Field}

To calculate the effective field for $U\neq 0$, we consider the case when
the dot levels lie within the conduction band. Then, scattering processes on a scale $D$
can involve real charge fluctuations of the dot.
Taking into account this effect and employing second-order perturbation theory in $V$, 
we continuously reduce the bandwidth $D$ by a positive infinitesimal $\delta D$. 
As a consequence, the dot energy levels in the dot are renormalized\cite{Hal78} as 
\bes
\begin{align}
\veps_{0}' &= \veps_{0} -  \sum_{\sm} \frac{\rho_{\sm}(-D)\delta D V^2}{\veps_{d\sm}+D} \,, \\
\veps_{1\sm}' &= \veps_{1\sm} - \delta D V^2 \left[ \frac{\rho_{\bsm}(-D)}{\veps_{d\bsm}+U+D} + \frac{\rho_{\sm}(+D)}{D-\veps_{d\sm}} \right] \,, \\
\veps_{2}' &= \veps_{2} - \sum_{\sm} \frac{\rho_{\sm}(+D)\delta D V^2}{D - \veps_{d\sm} - U} \,,
\end{align}
\label{eq:edocc}
\eds
where $\veps_{0}$ denotes the energy of the empty state,
$\veps_{1\sm}$ the energy of a singly occupied state with spin $\sm$, 
and $\veps_{2}$ the energy of the doubly occupied state.
Since $\veps_{d\sm} = \veps_{1\sm} - \veps_0$ by definition, for $U \to \infty$
Eqs.~\eqref{eq:edocc} yield the scaling equation for the single-particle energy,
\beq
\frac{d\veps_{d\sm}}{d\ln D} = -\frac{\wG_t}{2} \left[1 - \sqrt{\alpha\calt_b}\left(2\cos(\phi_{\sm}) + \cos(\phi_{\bsm})\right)\frac{D}{D_0}\right] \,.
\label{eq:evoled}
\edq
By integrating out the band from $D_0$ to $D$, we then obtain the renormalized energy level
\begin{multline}
\wteps_{d\sm} = \veps_{d\sm} \\
+ \frac{\wG_t}{2} 
\left[ \ln \frac{D_0}{D} - \sqrt{\alpha\calt_b}\left(2\cos(\phi_{\sm}) + \cos(\phi_{\bsm})\right) \left(1 - \frac{D}{D_0}\right) \right] \,.
\label{eq:edren}
\end{multline}

From Eq.~\eqref{eq:edren}, we find that the total level splitting
$\Delta_Z \equiv \wteps_{d\up} - \wteps_{d\down}$ is given by
\beq
\Delta_Z = (\veps_{d\up} - \veps_{d\down})
+ \wG_t\sqrt{\alpha\calt_b}\sin(\phi_{AB})\sin(\phi_{SO})\left(1 - \frac{D}{D_0}\right) \,.
\label{eq:effzeeman}
\edq
from which the effective magnetic field results,
\beq
g\mu_B B_{\rm eff}=\frac{\wG_t}{2}\sqrt{\alpha\calt_b}\sin(\phi_{AB})\sin(\phi_{SO})
\left(1 - \frac{D}{D_0}\right)\,.
\edq
Since the scaling terminates at $D = \wD$ roughly given by $\wD \approx |\wteps_{d}|$,
using Eq.~\eqref{eq:effzeeman} the compensating field $B_c=-B_{\rm eff}$ is given by
\beq\label{eq:bcuinf}
g\mu_B B_c = -\frac{\wG_t}{2} \sqrt{\alpha\calt_b}\sin(\phi_{AB})\sin(\phi_{SO})\left(1 + \frac{\wteps_d}{D_0}\right) \,,
\edq
where $\wteps_d$ can be found from,
\beq
\wteps_d = \veps_d + \frac{\wG_t}{2} \ln \frac{D_0}{|\wteps_d|} - \frac{3\wG_t}{2} \sqrt{\alpha\calt_b}\cos(\phi_{AB})\cos(\phi_{SO}) \,.
\label{eq:iurdlevel}
\edq
In the case $|\wteps_d|\ll D_0$, Eq.~\eqref{eq:bcuinf}
predicts that the effective field is {\em half} the value for the noninteracting case.
As a consequence, strong interactions reduce the external field
needed to compensate $B_{\rm eff}$ [see Eq.~\eqref{eq:bcnonint}].
Notably, the functional dependence
of $B_c$ on $\phi_{AB}$ and $\phi_{SO}$ remains the same.

On the other hand, for the special point $2\veps_d + U = 0$, the effective level evolves as,
\beq
\frac{d\veps_{d\sm}}{d\ln D} = \wG_t \sqrt{\alpha\calt_b}\cos(\phi_{\sm}) \frac{D}{D_0} \,,
\edq
so that for the compensating field we have,
\beq
g\mu_B B_c \approx -\wG_t\sqrt{\alpha\calt_b}\sin(\phi_{AB})\sin(\phi_{SO}) \,.
\label{eq:fucfield}
\edq
In this case, at $B = B_c$ the renormalized energy level reads
\beq
\wteps_d = \veps_d - \wG_t\sqrt{\alpha\calt_b}\cos(\phi_{AB})\cos(\phi_{SO}) \,. 
\label{eq:furdlevel}
\edq
Equation~\eqref{eq:fucfield} agrees, except for the prefactor,
with the mean-field result
obtained earlier, see Eq.~\eqref{eq:bcmf}.

\subsection{Kondo Temperature}

The charge fluctuation is quenched at $D < \wD$ and only spin fluctuations thus plays a role
at lower energies. In order to describe these fluctuations,
we perform a Schrieffer-Wolff transformation and obtain the Kondo Hamiltonian given by,
\begin{multline}
\calh_{K} = \sum_{k,k'} \left[J_+ S_+ c_{sk'\down}^{\dag}c_{sk\up} + J_- S_- c_{sk'\up}^{\dag}c_{sk\down} \right. \\
\left. + J_{z\up} S_z c_{sk'\up}^{\dag}c_{sk\up} - J_{z\down} S_z c_{sk'\down}^{\dag}c_{sk\down} \right]
+ K \sum_{k,k'} \sum_{\sm} c_{sk'\sm}^{\dag}c_{sk\sm} \,,
\label{eq:sdhams}
\end{multline}
where
\bes
\begin{align}
J_{\pm,z\sm} &= \frac{V^2}{|\wteps_d|} + \frac{V^2}{U - |\wteps_d|} \,, \\
K &= \frac{1}{2}\left(\frac{V^2}{|\wteps_d|} - \frac{V^2}{U - |\wteps_d|}\right) \,,
\end{align}
\eds
with renormalized single-particle energy $\wteps_d$ and bandwidth $\wD$.

Using Poor Man's scaling,\cite{anderson70}
the scaling equations for the coupling constants can be then written as
\bes
\begin{align}
\frac{d J_{z\sm}}{d\ln D} &= -2\trho_0 J_{\pm}^2 \,, \\
\frac{d J_{\pm}}{d\ln D} &= -\trho_0 J_{\pm} \left(J_{z\up} + J_{z\down}\right) \,,
\end{align}
\label{eq:jkscaling}
\eds
where we have kept only up to zeroth-order terms in $D/D_0$.
Since Eqs.~\eqref{eq:jkscaling} break down at $T_K$, we obtain
\beq
k_B T_K \approx \wD \exp\left[-\frac{1}{2\trho_0 J}\right] \,.
\edq
Here, we assume that
the spin splitting has been completely compensated
by an external magnetic field, thus employing
the renormalized single-particle energy at the compensating field 
given by Eq.~\eqref{eq:iurdlevel}.
For $U \to \infty$, the Kondo temperature can then be expressed
as a function of $\phi_{AB}$ and $\phi_{SO}$:
\beq
k_B T_K \approx T_K\left(\pi/2\right) \exp\left[-\frac{3}{2}\sqrt{\alpha\calt_b}\cos(\phi_{AB})\cos(\phi_{SO})\right] \,,
\label{eq:ktiu}
\edq
where $T_K(\pi/2)$ denotes the Kondo temperature at $\phi_{AB} = \pi/2$ and $\phi_{SO} = \pi/2$.
Note that the Kondo temperature is also a sinusoidal function of $\phi_{AB}$ and $\phi_{SO}$.
On the other hand, for the case $\varepsilon_d=-U/2$ we have
\beq
k_B T_K \approx T_K(\pi/2) \exp\left[ -\frac{\wG_t}{2\veps_d} \left(\sqrt{\alpha\calt_b}\cos(\phi_{AB})\cos(\phi_{SO})\right)^2 \right] \,.
\edq
In this case, the flux dependence is much weaker than that of the infinite $U$ case described by Eq.~\eqref{eq:ktiu}.

\section{Effective Field} \label{sec:Beff}

The qualitative discussion in Sec.~IV.B.
demonstrates the existence of an effective
field acting on the dot. To gain deeper insight into the properties
of the compensating field and investigate its functional
dependence on temperature and the gate voltage,
we now derive an effective Hamiltonian $\calh_{\rm eff}$
using perturbation theory.
Physically, the split Kondo peak can be understood
in terms of the dot valence instability
(virtual charge fluctuation) and spin-dependent density of states.
To deal with this instability, we perform
a Schrieffer--Wolff-like transformation
of the Hamiltonian given by Eq.~\eqref{eq:effham}.
Then,
\begin{multline}
\calh_{\rm eff} = \sum_{k,k'} \left[ \frac{V^2}{\veps_{d\up}-\veps_{k'\up}} c_{sk\up}c_{sk'\up}^{\dag} 
- \frac{V^2}{\veps_{d\down}-\veps_{k'\down}} c_{sk\down}c_{sk'\down}^{\dag} \right.
\\
\left.
+ \frac{V^2}{U + \veps_{d\up}-\veps_{k'\up}} c_{sk'\up}^{\dag}c_{sk\up}
- \frac{V^2}{U + \veps_{d\down}-\veps_{k'\down}} c_{sk'\down}^{\dag}c_{sk\down} \right] S_z \\
+ [\cdots] \,,
\end{multline}
where $[\cdots]$ includes spin-flip $S_{\pm}$ and potential scatterings.
At this point, unlike the usual  Schrieffer--Wolff transformation,
we employ a mean-field approximation for the lead electrons \cite{Kon04}
\bes
\begin{align}
\nbraket{c_{s k\sm}c_{s k'\sm'}^{\dag}} 
&=  \left[1 - f(\veps_{k\sm})\right] \delta_{k,k'}\delta_{\sm,\sm'} \,,
\\
\nbraket{c_{s k'\sm'}^{\dag}c_{s k\sm}} 
&=  f(\veps_{k\sm}) \delta_{k,k'}\delta_{\sm,\sm'} \,.
\end{align}
\eds
Then, the spin-flip scattering terms vanish and we obtain,
\begin{multline}
\calh_{\rm eff} = -V^2\int' d\omega~ \left[ \rho_{\up}(\omega) \frac{1-f(\omega)}{\omega - \veps_{d\up}}
- \rho_{\down}(\omega)\frac{1-f(\omega)}{\omega - \veps_{d\down}} \right. \\
\left. + \rho_{\up}(\omega)\frac{f(\omega)}{\omega - U - \veps_{d\up}}
- \rho_{\down}(\omega)\frac{f(\omega)}{\omega - U - \veps_{d\down}} \right] S_z
\\
+ [\text{Potential scattering}] \,.
\label{eq:exceff}
\end{multline}
Since the density of states is spin-dependent in our case,
the quantity in the square bracket is nonzero.

The effective Hamiltonian of Eq.~\eqref{eq:exceff} can be expressed as 
\beq
\calh_{\rm eff} = 2g\mu_B B_{\rm eff} S_z = g\mu_B B_{\rm eff} \sm_z \,,
\edq
from which it follows that
\begin{multline}
g\mu_B B_{\rm eff} = -\frac{\wG_t}{4}\int' d\omega~ \left\{ \right.\\ 
\left(1 + \sqrt{\alpha\calt_b}\cos(\phi_{\up})\frac{\omega}{D_0}\right)\left[\frac{1-f(\omega)}{\omega - \veps_{d\up}} + \frac{f(\omega)}{\omega - U - \veps_{d\up}} \right]
\\
\left.
- \left(1 + \sqrt{\alpha\calt_b}\cos(\phi_{\down})\frac{\omega}{D_0}\right)
\left[ \frac{1-f(\omega)}{\omega - \veps_{d\down}} + \frac{f(\omega)}{\omega - U - \veps_{d\down}} \right] \right\} \,.
\label{eq:exchange}
\end{multline}
This is the explicit formula of the \rm effective field and is a central result of our work.

For $B_{\rm ext}=0$ ($\veps_{d\up}=\veps_{d\down}$),
Eq.~\eqref{eq:exchange} can be written as
\begin{multline}
g\mu_B B_{\rm eff} = \wG_t\sqrt{\alpha\calt_b}\sin(\phi_{AB})\sin(\phi_{SO}) \\
\times \left\{1 - \frac{\veps_d}{2D_0} \wPsi(\veps_d)
+ \frac{(U + \veps_d)}{2D_0} \wPsi(U+\veps_d) \right\} \,,
\label{eq:Beff}
\end{multline}
where
\beq
\wPsi(\veps) = \ln \frac{2\pi}{\beta D_0} + \Re\left[\Psi\left(\frac{1}{2} -i\beta\frac{(\veps-E_F)}{2\pi} \right)\right] \,.
\edq
For low values of $\sqrt{\alpha \mathcal{T}_b}$, the effective field is proportional
to the lead polarization $P$ [cf. Eq.~\eqref{eq:polarization}], $B_{\rm eff}\propto \wG P$,
in analogy with a quantum dot coupled to ferromagnetic case.\cite{martinek}
The crucial difference is that in our case, $B_{\rm eff}$ is {\em nonvanishing}
for noninteracting electrons. Setting $U=0$ in Eq.~\eqref{eq:Beff}
we recover the expression found above.
Therefore, the effective magnetic field is not only generated via exchange interaction
between the dot electrons and the leads but it also contains a contribution
from the spin-orbit interaction and the magnetic flux when their associated
phases are nonzero at the same time.

As shown in the previous sections, the effective field $B_{\rm eff}$ can be compensated by applying an external magnetic field $B_c$ such that $B_c=-B_{\rm eff}$,\cite{note_self}
\begin{multline}\label{eq:bcsmall}
g\mu_B B_{c} = -\wG_t\sqrt{\alpha\calt_b}\sin(\phi_{AB})\sin(\phi_{SO}) \\
\times \left\{1 - \frac{\veps_d}{2D_0} \wPsi(\veps_d)
+ \frac{(U + \veps_d)}{2D_0} \wPsi(U+\veps_d) \right\} \,,
\end{multline}
which generalizes our previous expressions
[Eq.~\eqref{eq:bcuinf} and Eq.~\eqref{eq:fucfield}]
and is valid for nonzero temperature
and interacting electrons.
Importantly, the precise dependence on $\phi_{AB}$ and $\phi_{SO}$
remains in terms of periodic functions.

\begin{figure}
\centering
\includegraphics[width=0.4\textwidth]{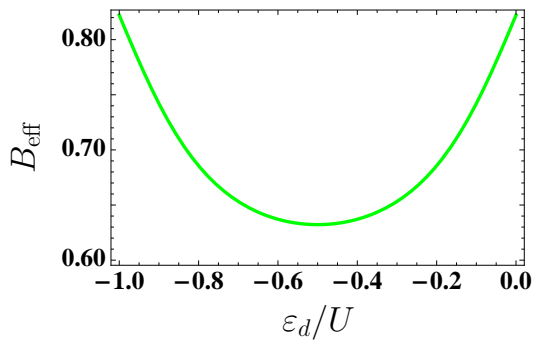}
\caption{Effective field $B_{\rm eff}$ as a function of $\veps_d/U$ with $U/D_0 = 1/2$ at $T=0$.
Here, the exchange field $B_{\rm eff}$ has been scaled
by $\wG_t\sqrt{\alpha\calt_b}\sin(\phi_{AB})\sin(\phi_{SO})$.
Refer to Eq.~\eqref{eq:Beff}.
}
\label{fig:BeffEd}
\end{figure}

In Fig.~\ref{fig:BeffEd} we show the effective field as a function of
the position of the quantum dot level, which can be tuned using a gate voltage.
The important result to bear in mind is that $B_{\rm eff}$ is nonzero
at the special point $2|\veps_d| + U =0$, due to the lack of particle-hole symmetry
in our system. This is in stark contrast with the case of ferromagnetic leads.
In that case, only the spin fluctuations prevail at the particle-hole symmetric point
so that the exchange field coming from charge fluctuations is zero. \cite{choi}

\section{Conclusions}\label{sec:concl}
To summarize, we have investigated the splitting that occurs
in the density of states of a quantum dot inserted in a mesoscopic interferometer
in the presence of spin-orbit interactions and magnetic flux.
In the Kondo regime, the resonance at the Fermi energy becomes
split at nonzero values of both the Aharonov-Bohm and the Rashba phases.
The splitting is due to an effective field whose main properties
can be more clearly derived from the instructive mapping to a Hamiltonian that
describes a quantum dot coupled to a transformed lead with spin-dependent
density of states. As a consequence, the coupling between the dot and the lead
depends on the spin orientation and an effective Zeeman splitting emerges.

For interacting electrons, a study of charge fluctuations within a scaling
procedure reveals an effective magnetic field that increases with
the charging energy. Importantly, the correction becomes of the same order
as the noninteracting value for $U\to\infty$.

The splitting can be compensated with an external magnetic field.
We have calculated the compensating field for both interacting and noninteracting
electrons. In both cases we obtain an expression which shows that the compensating
field is a periodic function of the Aharonov-Bohm and the spin-orbit phases.
We have also emphasized the breaking of particle-hole
symmetry in our system, which implies a nonzero value of the effective field
regardless of the applied gate voltage.


\section*{Acknowledgments}
This work was supported by the Spanish MICINN Grant No.\ FIS2008-00781,
the Conselleria d'Innovaci\'o, Interior i Justicia
(Govern de les Illes Balears, Spain),
and the Romanian National Research Program PN II-ID-502.


\begin{thebibliography}{90}
\bibitem{yac95}
A. Yacoby, M. Heiblum, D. Mahalu, and H. Shtrikman,
Phys. Rev. Lett. {\bf 74}, 4047 (1995).
\bibitem{lev95}
A. Levy Yeyati and M. B\"uttiker,
Phys. Rev. B {\bf 52}, R14360 (1995).
\bibitem{hac96}
G. Hackenbroich and H.A. Weidenm\"uller,
Phys. Rev. Lett. {\bf 76}, 110 (1996).
\bibitem{bru96}
C. Bruder, R. Fazio and H. Schoeller,
Phys. Rev. Lett. {\bf 76}, 114 (1996).
\bibitem{shu97}
R. Schuster, E. Buks, M. Heiblum, D. Mahalu,
V. Umansky and H. Shtrikman, Nature {\bf 385}, 417 (1997).
\bibitem{ji00}
Y. Ji, M. Heiblum, D. Sprinzak, D. Mahalu, and H. Shtrikman,
Science {\bf 290}, 779 (2000).
\bibitem{sig04}
M. Sigrist, A. Fuhrer, T. Ihn, K. Ensslin, S.E. Ulloa,
W. Wegscheider, and M. Bichler,
Phys. Rev. Lett. {\bf 93}, 066802 (2004).
\bibitem{kob02}
K. Kobayashi, H. Aikawa, S. Katsumoto, and Y. Iye,
Phys. Rev. Lett. \textbf{88}, 256806 (2002).


\bibitem{Hewson93}
A. C. Hewson, {\it The Kondo Problem to Heavy Fermions} (Cambridge University Press, Cambridge, 1993).

\bibitem{Kondo}
D.~Goldhaber-Gordon, H.~Shtrkman, D.~Mahalu, D.~Abusch-Magder, U.~Meirav,
M.~A.~Kastner, Nature {\bf 391}, 156 (1998); S.~M.~Cronenwett, T.~H.~Oosterkamp, and L.~P.~Kouwenhoven, Science {\bf 281}, 540 (1998);
J.~Schmid, J.~Weis, K.~Eberl, and K.~von Klitzing, Physica B {\bf 256}, 182 (1998).


\bibitem{hof01}
W. Hofstetter, J. K\"onig, and H. Schoeller,
Phys. Rev. Lett. \textbf{87}, 156803 (2001).
\bibitem{bul01}
B.R. Bulka and P. Stefa\'nski,
Phys. Rev. Lett. {\bf 86}, 5128 (2001).

\bibitem{fab07}
For a recent review, see, e.g, J. Fabian, A. Matos-Abiague, C. Ertler, P. Stano, and I. Zutic,
Acta Physica Slovaca {\bf 57}, 565 (2007).
\bibitem{Rashba60} E. I. Rashba,
Fiz. Tverd, Tela (Leningrad) {\bf 2}, 1224 (1960). [Sov. Phys. Solid State {\bf 2}, 1109 (1960)]

\bibitem{morpurgo}
A. F. Morpurgo, J. P. Heida, T. M. Klapwijk, B. J. van Wees, and G. Borghs,
Phys. Rev. Lett. {\bf 80}, 1050 (1998).
\bibitem{yau}
J. B. Yau, E. P. De Poortere, and M. Shayegan,
Phys. Rev. Lett. {\bf 88}, 146801 (2002).
\bibitem{yang}
M. J. Yang, C. H. Yang, and Y. B. Lyanda-Geller,
Europhys. Lett. {\bf 66}, 826 (2004).
\bibitem{grbic}
B. Grbi\'c, R. Leturcq, T. Ihn, K. Ensslin, D. Reuter, and A. D. Wieck,
Phys. Rev. Lett. {\bf 99}, 176803 (2007).
\bibitem{sun05}
Q.F. Sun, J. Wang, and H. Guo, Phys. Rev. B {\bf 71}, 165310 (2005);
Q.-f.Sung and X.C. Xie, Phys. Rev. B {\bf 73}, 235301 (2006).
\bibitem{lop07}
R. L\'opez, D. S\'anchez, and Ll. Serra, Phys. Rev. B {\bf 76}, 035307 (2007).
\bibitem{cri09}
M. Crisan, D. S\'anchez, R. L\'opez, Ll. Serra, and I. Grosu,
Phys. Rev. B {\bf 79}, 125319 (2009).
\bibitem{hea08}
R.J. Heary, J.E. Han, and L. Zhu, Phys. Rev. B {\bf 77}, 115132 (2008).
\bibitem{ver09}
E. Vernek, N. Sandler, and S. E. Ulloa,
Phys. Rev. B {\bf 80}, 041302(R) (2009).

\bibitem{martinek}
J. Martinek, Y. Utsumi, H. Imamura, J. Barna\'s, S. Maekawa, J. K\"onig, and G. Sch\"on,
Phys. Rev. Lett. {\bf 91}, 127203 (2003);
J. Martinek, M. Sindel, L. Borda, J. Barna\'s, J. K\"onig, G. Sch\"on, and J. von Delft,
Phys. Rev. Lett. {\bf 91}, 247202 (2003).
\bibitem{choi}
R. L\'opez and D. S\'anchez,
Phys. Rev. Lett. \textbf{90}, 116602 (2003);
M.-S. Choi, D. S\'anchez, and R. L\'opez,
Phys. Rev. Lett. \textbf{92}, 056601 (2004).
\bibitem{krawiec}
M. Krawiec, J. Phys.: Condens. Matter {\bf 19}, 346234 (2007).
\bibitem{zeeman}
We have neglected the Zeeman interaction due to the
applied magnetic flux since in experimentally available
mesoscopic interferometers the applied fields are usually
quite small (of the order of mT).\cite{morpurgo,grbic}
\bibitem{anderson61}
P. W. Anderson, Phys. Rev. {\bf 124}, 41 (1961).
\bibitem{hor}
B. Horvati\'c, D. Sokcevi\'c, and V. Zlati\'c,
Phys. Rev. B {\bf 36}, 675 (1987).
\bibitem{wilson}
K. G. Wilson, Rev. Mod. Phys. {\bf 47}, 773 (1975).
\bibitem{Kri80}
H. R. Krishna-murthy, J. W. Wilkins, and K. G. Wilson, Phys. Rev. B {\bf 21}, 1003 (1980).
\bibitem{Hof00} 
W. Hofstetter, Phys. Rev. Lett. {\bf 85}, 1508.
\bibitem{Wei07} 
R. Peters, T. Pruschke, and F. B. Anders, Phys. Rev. B {\bf 74}, 245114 (2006);
A. Weichselbaum and J. von Delft, Phys. Rev. Lett. {\bf 99}, 076402 (2007).



\bibitem{yos08}
R. Yoshii and M. Eto,
J. Phys. Soc. Jap. {\bf 77}, 123714 (2008).

\bibitem{Hal78} 
F. D. M. Haldane, Phys. Rev. B. {\bf 40}, 416 (1978).
\bibitem{anderson70}
P. W. Anderson, Phys. Rev. {\bf 3}, 2436 (1970).
\bibitem{Kon04}
J. K\"onig, J. Martinek, J. Barna\'s, and G. Sch\"on, in {\em CFN Lectures on Functional Nanostructures},
Eds. K. Busch et al., Lecture Notes in Physics {\bf 658}, Springer, pp. 145-164 (2005). 
\bibitem{note_self}
We emphasize, however, that $B_{\rm eff}$
depends on $B_{\rm ext}$, see the denominators in Eq.~\eqref{eq:exchange}.
As a consequence, the condition $B_c=-B_{\rm eff}$ must be solved {\em self-consistently}.
Nevertheless, for the problem we consider here we have checked that including
self-consistency merely introduces unimportant corrections to the value
predicted by Eq.~\eqref{eq:bcsmall}.
\end{thebibliography}
\end{document}